\documentclass[11pt]{amsart}

\usepackage[T1]{fontenc}
\usepackage[utf8]{inputenc}
\usepackage{lmodern}
\usepackage{amsmath,amssymb,amsthm,mathtools}
\usepackage{graphicx}
\usepackage{booktabs}
\usepackage{multirow}
\usepackage{array}
\usepackage{enumitem}
\usepackage{float}
\usepackage{hyperref}
\usepackage[nameinlink,noabbrev]{cleveref}
\usepackage[margin=1in]{geometry}
\usepackage{setspace}
\usepackage{xcolor}
\usepackage{url}
\usepackage{tikz}
\usetikzlibrary{arrows.meta,positioning}
\usepackage[
  backend=biber,
  style=numeric,
  sorting=none,
  giveninits=false,
  doi=true,
  isbn=true,
  url=true,
  eprint=true
]{biblatex}

\hypersetup{
    colorlinks=true,
    linkcolor=blue,
    citecolor=blue,
    urlcolor=blue
}

\title[GlobalCY I: Globally Defined and Symmetry-Aware Neural Kähler Potentials]{GlobalCY I: A JAX Framework for Globally Defined and Symmetry-Aware Neural Kähler Potentials}
\author{Abdul Rahman}
\thanks{Email: arahman@alum.howard.edu}

\begin{document}
\maketitle

\begin{abstract}
We present \emph{GlobalCY}, a JAX-based framework for globally defined and symmetry-aware neural K\"ahler-potential models on projective hypersurface Calabi--Yau geometries. The central problem is that local-input neural K\"ahler-potential models can train successfully while still failing the geometry-sensitive diagnostics that matter in hard quartic regimes, especially near singular and near-singular members of the Cefal\'u family. To study this, we compare three model families---a local-input baseline, a globally defined invariant model, and a symmetry-aware global model---on the hard Cefal\'u cases $\lambda=0.75$ and $\lambda=1.0$ using a fixed multi-seed protocol and a geometry-aware diagnostic suite. In this benchmark, the globally defined invariant model is the strongest overall family, outperforming the local baseline on the two clearest geometric comparison metrics, negative-eigenvalue frequency and projective-invariance drift, in both cases. The gains are strongest at $\lambda=0.75$, while $\lambda=1.0$ remains more difficult. The current symmetry-aware model improves projective-invariance drift relative to the local baseline, but does not yet surpass the plain global invariant model. These results show that global invariant structure is a meaningful architectural constraint for learned K\"ahler-potential modeling in hard quartic Calabi--Yau settings.
\end{abstract}

\section{Introduction}

The computation of explicit Ricci-flat K\"ahler metrics on Calabi--Yau manifolds remains one of the central computational problems at the interface of complex differential geometry, string compactification, and scientific machine learning. Yau's solution of the Calabi conjecture guarantees existence and uniqueness of a Ricci-flat metric in a fixed K\"ahler class, but in general does not provide a constructive closed-form formula. In practice, this has led to a long-standing computational program built from numerical differential geometry, structured finite-dimensional ans\"atze, and, more recently, machine-learned surrogates for K\"ahler potentials and associated metric data.

In recent years, machine learning has emerged as a promising tool for this problem. Neural models can, in principle, approximate complicated geometric quantities without requiring a fully explicit analytic solution. However, the central challenge in this setting is not merely approximation power. A learned K\"ahler-potential model is valuable only to the extent that it respects the geometric structure of the underlying manifold. A model may optimize a training objective successfully and still fail in precisely the ways that matter most scientifically: instability in hard regimes, poor behavior under projective rescaling, loss of geometric consistency across views, or degradation on curvature- and topology-sensitive diagnostics. In this sense, the problem is not only one of regression, but one of geometric fidelity.

This issue becomes especially acute in singular and near-singular quartic families. Recent work, particularly that of Berglund et al., shows that in the Cefal\'u setting local-input $\Phi$-style models can exhibit weaknesses that are not obvious from loss values alone, while more global or spectral constructions behave more robustly on geometry-sensitive quantities \cite{BerglundEtAl2023}. That observation changes the emphasis of the problem. The main bottleneck is not simply whether a neural model can fit local data, but whether the learned ansatz carries enough global geometric structure to remain faithful in difficult regimes.

The present paper takes that architectural question as its main object of study. Rather than treating globality and symmetry-awareness as secondary refinements, we make them explicit modeling principles and test their effect in a controlled computational setting. Concretely, we build and compare three model families for learned K\"ahler-potential corrections: a local-input baseline, a globally defined invariant model, and a symmetry-aware global model. Our focus is on hard quartic regimes in which geometric fragility is already known to appear, namely the Cefal\'u cases $\lambda=0.75$ and $\lambda=1.0$.

The scientific aim of the paper is modest but precise. We do not claim to solve the full Ricci-flat metric problem for Calabi--Yau manifolds, nor do we claim that a first symmetry-aware implementation should already be optimal. Instead, we ask a narrower question: does imposing global invariant structure improve the behavior of learned K\"ahler-potential models on the geometry-sensitive diagnostics that matter most in hard quartic regimes? Our thesis is that it does. In particular, we argue that global invariant structure is not a cosmetic modeling choice, but a scientifically meaningful architectural constraint.

This question matters for two reasons. First, hard quartic regimes provide a stringent stress test for learned geometric models: if an architectural improvement is real, it should become visible there before it becomes visible in easier cases. Second, the field needs reusable computational infrastructure that supports controlled scientific comparison rather than one-off experiments. To that end, the present work is designed not only as a model-comparison study, but also as a reproducible framework in which geometry generation, model evaluation, diagnostic computation, and reproducible result export are cleanly separated and systematically organized.

\paragraph{\textbf{Relation to earlier work}}
Earlier work showed that not all learned metric models behave equally well: some can achieve acceptable training behavior while still giving unstable geometric quantities on hard examples. In particular, Berglund et al. highlighted that more global constructions may behave better than purely local-input models \cite{BerglundEtAl2023}. The present paper develops that idea into a full comparison framework. Rather than asking only whether a neural model can be trained, we ask whether giving the model the right global and symmetry-aware structure makes it behave more like the geometric object it is supposed to represent.

\paragraph{\textbf{Motivation}}
The motivation for GlobalCY is not merely to provide another implementation of a learned metric ansatz, but to make a specific class of geometry-first experiments possible in a reusable way. Existing work has already shown that local-input neural K\"ahler-potential models can look acceptable under ordinary optimization criteria while still behaving poorly on geometry-sensitive diagnostics in hard quartic regimes. What is needed, therefore, is not just more training capacity, but a computational framework in which global invariant structure, symmetry-aware inputs, differentiable metric construction, and reproducible model comparison can all be studied together. GlobalCY is designed to supply exactly that. By building the local, globally invariant, and symmetry-aware model families into a shared JAX-native environment, and by coupling them to reproducible GeoCYData bundles and frozen result workflows, the framework makes it easier to test architectural ideas, diagnose geometric failure modes, and carry the strongest models forward into later work on symbolic distillation, moduli dependence, singularity-aware learning, and eventually downstream physical observables.

\paragraph{\textbf{Why JAX}}
The learned K\"ahler-potential framework studied here requires repeated computation of derivatives of \(\phi\), construction of Hermitian metric corrections of the form \(g = g_{\mathrm{FS}} + \partial \bar{\partial}\phi\), and reproducible evaluation of geometry-aware diagnostics across multiple model families and hard parameter regimes. JAX is particularly well suited to this setting because it combines NumPy-style array programming with automatic differentiation, compilation, batching, and accelerator portability in a single scientific-computing environment \cite{BradburyEtAlJAX,HeekEtAlFlax}. This allows the geometric construction, the learned model, and the diagnostic layer to live in the same differentiable framework rather than being coupled only through post hoc numerical wrappers. For the present paper, this is not merely a convenience: it is part of what makes the framework inspectable, reproducible, and extensible enough to support later work on symbolic distillation, moduli dependence, and singularity-aware model classes.

\paragraph{\textbf{Contributions}}
The main contributions are as follows.
\begin{enumerate}[leftmargin=2em]
    \item We introduce \emph{GlobalCY}, a JAX-native framework for globally defined and symmetry-aware neural K\"ahler-potential models on projective hypersurface Calabi--Yau geometries.
    
    \item We implement a controlled architectural comparison among three model families: a local-input baseline, a globally defined invariant model, and a symmetry-aware global model.
    
    \item We evaluate these models on two hard Cefal\'u quartic regimes, $\lambda=0.75$ and $\lambda=1.0$, using a fixed multi-seed protocol designed to isolate model-class effects from preprocessing or case-definition confounds.
    
    \item We assess model behavior with a geometry-aware diagnostic suite that includes negative-eigenvalue frequency, minimum-eigenvalue behavior, projective-invariance drift, symmetry consistency, and related summary statistics, thereby emphasizing geometric fidelity rather than training loss alone.
    
    \item We provide a reproducible result layer, including comparison tables, figures, and frozen summary artifacts, so that the computational benchmark can be carried directly into scientific interpretation and manuscript construction.
\end{enumerate}

\paragraph{\textbf{Code and research organization}}
The research program surrounding this paper is maintained through the \texttt{geocy-labs} GitHub organization, \url{https://github.com/geocy-labs}, which serves as a public home for ongoing work at the intersection of geometry-aware machine learning, computational Calabi--Yau geometry, and scientific software infrastructure. The \emph{GlobalCY} framework introduced here is available at \url{https://github.com/geocy-labs/globalcy}, and the benchmark data and geometric substrate used in the present experiments are provided through \texttt{GeoCYData} at \url{https://github.com/geocy-labs/geo-cy-data}. Together, these repositories expose the organizational, geometric, and model-layer components underlying the controlled comparisons reported in this paper.
\section{Related Work}

\subsection{Classical numerical approaches to Ricci-flat K\"ahler metrics}

The modern computational study of Ricci-flat K\"ahler metrics on Calabi--Yau manifolds begins with the foundational work of Calabi and Yau. Calabi formulated the problem of prescribing Ricci curvature within a fixed K\"ahler class, and Yau's proof of the Calabi conjecture established that every compact K\"ahler manifold with vanishing first Chern class admits a unique Ricci-flat K\"ahler metric in each K\"ahler class \cite{Calabi1954,Yau1978}. This theorem provides the conceptual basis for all later numerical and machine-learning approaches: the target metric exists canonically, but in general is not available in explicit closed form.

The first sustained computational approaches to this problem emerged from numerical differential geometry and finite-dimensional approximation theory. Donaldson's work on balanced metrics and projective approximations showed that Ricci-flat metric computation can be approached through structured algebraic ans\"atze rather than brute-force discretization alone \cite{Donaldson2005,Donaldson2009}. A central lesson of this line of work is that approximation quality must be judged by intrinsically geometric criteria, not merely by numerical convenience.

Douglas, Karp, L\"ukic, and Reinbacher brought this program into explicit Calabi--Yau settings by constructing numerical approximations to Calabi--Yau metrics on nontrivial examples \cite{DouglasKarpLukicReinbacher2008}. Their work made clear that the metric problem is computationally accessible in practice, provided one combines suitable sampling, a geometrically meaningful ansatz, and diagnostics that probe the quality of the resulting metric as a geometric object.

These classical approaches remain essential background for the present paper. They established three principles that continue to structure the subject: first, that the metric problem is naturally approached through a K\"ahler-potential or algebraic ansatz; second, that geometric fidelity matters more than raw optimization success; and third, that the amount of global structure built into the approximation is a scientifically consequential modeling choice.

\subsection{Machine learning approaches to Calabi--Yau metrics}

Machine learning entered the subject as a natural extension of these earlier numerical ideas. Ashmore, He, and Ovrut showed that neural methods can be used to approximate Calabi--Yau metrics and related quantities while remaining meaningfully tied to the underlying geometric problem \cite{AshmoreHeOvrut2020}. Their work helped establish that machine learning could serve as a serious computational tool in this setting rather than merely as an interpolation heuristic.

Subsequent work broadened the model classes used for learned metric approximation. In particular, Douglas et al.\ developed holomorphic-network-based constructions for numerical Calabi--Yau metrics, illustrating that learned ans\"atze can be designed so as to preserve more of the complex-geometric structure of the problem than generic black-box regressors \cite{DouglasEtAl2022}. Around the same period, the \emph{cymetric} line of work emphasized practical neural pipelines for learning Calabi--Yau metrics and related geometric quantities on broader classes of examples \cite{LarforsSchneiderRuehle2021}. Taken together, these developments reinforced a lesson that is now central to the field: in learned Calabi--Yau geometry, the quality of the ansatz matters as much as the expressive power of the optimizer.

The closest direct precursor to the present paper is Berglund et al., who study machine-learned Calabi--Yau metrics and curvature on smooth and singular families, with particular emphasis on difficult quartic geometries including the Cefal\'u pencil \cite{BerglundEtAl2023}. Their work is especially relevant for two reasons. First, it shows that topology- and curvature-sensitive quantities reveal failure modes that are invisible at the level of training loss alone. Second, it provides evidence that more global or spectral constructions can outperform local-input $\Phi$-style models in singular and near-singular quartic regimes. The present paper takes this lesson as its starting point, but narrows the question to a more controlled architectural comparison: whether globally defined invariant structure and symmetry-aware structure improve learned K\"ahler-potential behavior on hard quartic cases.

\subsection{Geometry-aware software and architectural structure}

A parallel line of work has emphasized that progress in this area depends not only on model design, but also on reusable geometry-aware computational infrastructure. Gerdes and Krippendorf's \emph{CYJAX} is an important example: it provides a JAX-native package for machine learning Calabi--Yau metrics and demonstrates the value of modular, differentiable, geometry-aware software for reproducible experimentation \cite{GerdesKrippendorf2022}. The present work shares this software-oriented emphasis, but its primary focus is narrower: not general package design alone, but controlled comparison of learned K\"ahler-potential architectures on a fixed hard-regime benchmark. In particular, CYJAX is centered on an algebraic ansatz for the K\"ahler potential with built-in K\"ahlerity and patch-overlap compatibility, whereas the present work isolates a different question: whether globally invariant and symmetry-aware learned representations improve geometric fidelity relative to local-input baselines on hard quartic regimes.

Recent work has also sharpened the importance of architectural constraints that encode global or group-theoretic structure. Hendi, Larfors, and Walden develop group-invariant Calabi--Yau metric learning based on canonicalization to a fundamental domain and explicit discrete-symmetry reduction \cite{HendiLarforsWalden2024}. Their results show that symmetry can be used as an active modeling ingredient rather than treated as a passive property of the underlying manifold. This is closely aligned with one of the main themes of the present paper. Our emphasis, however, is somewhat different. Rather than studying symmetry reduction in isolation, we compare a three-level architectural ladder---local-input, globally invariant, and symmetry-aware global models---on hard quartic cases where geometric fragility is already known to appear.

Ek, Kim, and Mishra pursue a related but distinct direction through Grassmannian learning combined with Donaldson-style approximation ideas \cite{EkKimMishra2024}. Their framework is especially valuable because it pushes the subject toward principled geometric ans\"atze with built-in positivity and structured approximation, rather than relying on unconstrained black-box learning. In a different language, it raises the same foundational question that motivates the present work: how much of the geometric burden should be carried by the architecture itself rather than recovered imperfectly through optimization?

\subsection{Toward downstream geometry and physics}

A further important development is the recent expansion of the computational stack beyond metric learning alone. The \emph{cymyc} framework extends learned and numerical Calabi--Yau computation toward tensor fields of higher degree, curvature quantities, Yukawa couplings, and related downstream calculations \cite{ButbaiaEtAl2025}. This development is important for positioning the present paper. This paper is motivated by the broader goal of building learned metric models that are stable and interpretable enough to support more ambitious geometry-sensitive and physics-facing computations later.

For that reason, the present work should be read as part of a larger transition in the field: from proof-of-concept metric surrogates toward reusable computational infrastructure capable of supporting broader scientific workflows. In that transition, the architectural question studied here is foundational. If the learned K\"ahler-potential model is not globally well behaved in hard regimes, then downstream curvature, characteristic-form, or observable-level computations will inherit that instability.

\subsection{Interpretability and moduli-dependent directions}

If learned Calabi--Yau metric models are to become scientifically useful beyond benchmark comparison, two further directions become unavoidable: interpretability and moduli dependence. On the interpretability side, recent work has begun to connect high-performing neural surrogates to symbolic or analytic approximations. Mirjani\'c and Mishra study symbolic approximations to Ricci-flat metrics using extrinsic symmetries of Calabi--Yau hypersurfaces and show that, in favorable situations, these symmetries can lead to compact analytic representations and improved neural constructions \cite{MirjanicMishra2024}. More recently, Eng develops an explicitly symbolic-distillation approach in which neural approximations are compressed into compact analytic formulas while preserving high predictive accuracy on the studied examples \cite{Eng2026}. These works suggest a plausible route from black-box metric surrogates to interpretable formulas that can be analyzed and reused mathematically.

On the moduli-dependent side, the next challenge is to move beyond isolated benchmark points and learn metric families that vary coherently with geometric parameters. Constantin, Lukas, and Nutricati develop approximate Ricci-flat metrics with explicit K\"ahler-moduli dependence by combining numerical learning with symbolic regression \cite{ConstantinLukasNutricati2026}. This direction is directly relevant to the longer-term program motivating the present paper. This work deliberately restricts attention to two hard Cefal\'u cases in order to isolate the architectural question cleanly. But one of its broader aims is to establish a framework in which globally defined and symmetry-aware model classes can later be transported across moduli rather than trained case-by-case in isolation.

Taken together, these developments clarify the niche of the present paper. We do not attempt to solve all aspects of machine-learned Calabi--Yau geometry at once. Instead, we focus on what we view as a necessary first step: establishing, in a controlled and reproducible way, that globally defined invariant structure is a better architectural starting point than local-input baselines on hard quartic regimes. Stronger symmetry-aware gains, symbolic compression, moduli dependence, and downstream geometry or physics calculations are then natural next layers of the program rather than claims that must be forced into the first paper.

\section{Problem Statement and Thesis}

\subsection{Problem statement}

A central difficulty in learned Calabi--Yau metric approximation is that numerical success and geometric fidelity are not the same thing. Local-input neural K\"ahler-potential models may fit their training objectives reasonably well while still failing the geometric tests that matter most: compatibility with global projective structure, stability under coordinate changes, and robustness of geometry-sensitive diagnostics in difficult regimes. This issue becomes especially acute near singular and near-singular members of the Cefal\'u family, where architectural weaknesses that remain hidden in easier cases become visible.

The problem addressed in this paper is therefore not merely whether a neural network can be trained to represent a scalar K\"ahler-potential correction. The real question is whether the learned ansatz behaves like a geometrically meaningful object once it is used to construct a metric. A model may appear successful at the level of optimization while still being globally incorrect in ways that matter scientifically. In particular, patch-local constructions may exhibit acceptable loss values yet fail under projective rescaling, produce unstable spectral behavior, or degrade on diagnostics that probe geometric consistency rather than raw predictive fit.

This shifts the bottleneck from expressivity alone to architectural structure. The relevant issue is not simply whether the model is flexible enough, but whether it is given enough global geometric information to support stable and faithful computation in the hard quartic regimes where failure is most informative. The present paper studies this bottleneck on a deliberately controlled first testbed: the hard Cefal\'u cases $\lambda=0.75$ and $\lambda=1.0$. The guiding question is whether globally defined and symmetry-aware neural K\"ahler-potential models provide more faithful learned metric corrections than local-input baselines in these difficult regimes.

\subsection{Thesis}

The thesis of this paper is that globality and symmetry-awareness are not cosmetic additions to a learned K\"ahler-potential model. They are scientifically meaningful architectural constraints, and imposing them improves the geometric behavior of the resulting learned metric corrections on hard quartic Calabi--Yau regimes.

More specifically, we argue that learned K\"ahler-potential modeling should be treated as a geometry-first problem. In this setting, the ansatz is not secondary to the optimizer; it is one of the principal determinants of whether the learned object behaves like a genuine geometric correction or merely like a locally trained numerical surrogate. The main claim advanced here is correspondingly narrow but substantive: among the model families compared in this first study, the globally defined invariant model is the strongest overall on the clearest geometry-sensitive diagnostics. This supports the view that global invariant structure is already a useful and nontrivial architectural principle for hard-regime learned metric approximation.

\subsection{Hypothesis}

Our working hypothesis is that globally defined invariant models, and eventually stronger symmetry-aware models, should improve learned metric behavior on the diagnostics that most directly probe geometric fidelity. In the present study, the most important expected improvements are:
\begin{itemize}[leftmargin=2em]
    \item lower negative-eigenvalue frequency,
    \item lower projective-invariance drift,
    \item improved geometric consistency, and
    \item greater robustness in singular and near-singular quartic regimes.
\end{itemize}

We test this hypothesis through a controlled multi-seed comparison of three model families: a local-input baseline, a globally defined invariant model, and a symmetry-aware global model. The expected outcome is not that every diagnostic must improve simultaneously, nor that a first symmetry-aware implementation must already dominate. Rather, the expectation is that global invariant structure should already yield measurable gains on the clearest comparison metrics, while the harder regime should reveal where current architectures remain insufficient.

\subsection{Assumptions}

We work within the standard K\"ahler-potential correction ansatz
\[
g = g_{\mathrm{FS}} + \partial \bar{\partial}\phi,
\]
and adopt the following assumptions:
\begin{enumerate}[leftmargin=2em]
    \item the hard quartic Cefal\'u cases provide an appropriate first stress test for architectural comparison;
    \item geometry-aware diagnostics are more informative than training loss alone for assessing learned metric quality; and
    \item global invariant structure and symmetry-aware structure are plausible sources of improved numerical behavior.
\end{enumerate}

These assumptions are intentionally modest. We do not assume that the present models solve the full Ricci-flat metric problem, nor that the current symmetry-aware implementation is already optimal. We assume only that hard quartic regimes provide a meaningful setting in which architectural differences become visible, that learned metric models should be judged by geometry-facing diagnostics rather than optimization success alone, and that incorporating global structure is a principled place to seek the first real improvements.

Under these assumptions, the present paper should be read as a controlled first step rather than a final resolution of the problem. Its role is to isolate one architectural claim clearly: that global invariant structure already improves learned K\"ahler-potential behavior on a hard quartic benchmark. Stronger symmetry-aware modeling, symbolic distillation, moduli-dependent learning, and singularity-aware refinement are then natural next stages of the broader program rather than claims that need to be compressed prematurely into this first paper.

\section{Geometric Setup}

\subsection{The quartic hypersurface setting}

We work with smooth and nearly singular quartic hypersurfaces in projective space, viewed as compact K\"ahler manifolds equipped with the metric ansatz
\[
g = g_{\mathrm{FS}} + \partial \bar{\partial}\phi,
\]
where \(g_{\mathrm{FS}}\) is the pullback of the ambient Fubini--Study metric and \(\phi\) is the learned scalar correction. This setting is computationally attractive for several reasons. It provides a concrete projective realization in which homogeneous coordinates, local charts, and projective-invariant feature constructions arise naturally. At the same time, it is rich enough to exhibit nontrivial geometric difficulty, especially in singular and near-singular regimes. It also aligns well with prior work on learned Calabi--Yau and K3 metric approximation, making controlled comparison with the existing literature possible \cite{BerglundEtAl2023,GerdesKrippendorf2022}.

For the purposes of this paper, quartic hypersurfaces provide a controlled arena in which local-input, globally invariant, and symmetry-aware K\"ahler-potential models can be compared under a common protocol. The point is not to survey all Calabi--Yau families at once, but to isolate a compact and scientifically meaningful geometry class in which the difference between local and global architectural structure is already visible. In this setting, projective invariance, chart dependence, and symmetry constraints can be imposed and tested concretely at the level of the learned ansatz rather than discussed only abstractly.

This choice also matches the current computational stack. GeoCYData supplies reproducible quartic-family bundles with local chart views, invariant feature views, case metadata, and, when available, symmetry-aware metadata. GlobalCY consumes these bundles to construct, train, and compare learned K\"ahler-potential models under fixed cases and seeds. The quartic hypersurface setting is therefore not only mathematically natural, but also the regime in which the present geometry-to-model pipeline is most mature.

\subsection{The Fermat quartic}

The Fermat quartic serves as the simplest reference geometry in the present program. Because it is highly symmetric and comparatively benign, it provides a useful scaffold on which software components, bundle interfaces, model classes, and diagnostic code paths can be validated before moving to the harder Cefal\'u regimes. In practice, it has functioned as the smoke-test geometry for the initial GlobalCY implementation, allowing the local-input and globally invariant pipelines to be checked in a stable setting before harder-regime comparisons are attempted.

Its role in the present paper is therefore methodological rather than central. The Fermat quartic is not one of the main comparison cases in the final benchmark, and the paper does not draw its substantive scientific conclusions from it. Instead, it serves as a clean reference geometry for separating implementation issues from hard-regime geometric effects. By validating the framework first on a comparatively well-behaved quartic, one gains confidence that later failures in the Cefal\'u family arise from geometry and model structure rather than from basic defects in the computational pipeline.

\subsection{The Cefal\'u family}

The main geometric testbed in this study is the Cefal\'u quartic family, which has already emerged in prior work as a particularly revealing setting for learned metric approximation in hard quartic regimes \cite{BerglundEtAl2023}. Its importance lies in the presence of singular and near-singular members whose geometry is sufficiently delicate to expose weaknesses that remain hidden in easier smooth cases. In such regimes, the relevant question is not merely whether a model trains, but whether architectural improvements translate into better behavior on geometry-sensitive diagnostics.

Within this family, we focus on the two cases
\[
\lambda = 0.75, \qquad \lambda = 1.0.
\]
These values were chosen because they provide a meaningful first hard-regime comparison while keeping the scope of the study controlled and reproducible. Both lie in a regime where quartic behavior becomes more delicate, and together they form a natural first pair for testing whether globally invariant and symmetry-aware modeling improve learned metric behavior relative to local-input baselines.

From the standpoint of the software stack, these are also the cases for which the current pipeline is most complete. GeoCYData provides stable case support, canonicalized identifiers, symmetry-aware metadata, and multi-seed benchmark inputs, while GlobalCY provides smoke-scale validation, multiseed comparison workflows, and the result-export layer built around these cases. The Cefal\'u family is therefore not only the correct scientific focus for this first benchmark, but also the geometric core around which the present computational evidence is organized.

\subsection{Why the hard regime matters}

The emphasis on the selected Cefal\'u cases is motivated by the fact that they are already known to be geometrically difficult. Prior work indicates that singular and near-singular quartic regimes are precisely where topology- and curvature-sensitive diagnostics begin to separate competing learned metric ans\"atze in a meaningful way \cite{BerglundEtAl2023}. In easier settings, several model classes may appear acceptable. In harder regimes, by contrast, structural deficiencies become visible: local-input models may degrade, sensitive quantities become unstable, and the difference between a merely predictive surrogate and a genuinely geometric one becomes much sharper.

This motivates the deliberately narrow scope of the present study. We do not attempt to benchmark every quartic case or cover the entire Cefal\'u family in one paper. Instead, we focus on two scientifically informative hard cases and compare model families under a fixed multi-seed protocol. This makes the resulting claim sharper. If global invariant structure is genuinely useful, then its advantage should already become visible in the regimes where naive local surrogates are under the greatest stress.

The results support that choice of testbed. Across both selected cases, the globally defined invariant model is strongest on the clearest geometry-sensitive comparison metrics, while the \(\lambda=1.0\) regime remains more difficult in the sense that the separation from the local baseline is smaller there. Thus, these Cefal\'u cases are not merely benchmark labels. They identify the first region of the quartic landscape in which the architectural question studied here becomes scientifically visible.
\section{The GlobalCY Framework}

\subsection{Software architecture}

\emph{GlobalCY} is a JAX-native framework for learned K\"ahler-potential modeling on projective hypersurface Calabi--Yau geometries, with particular emphasis on globally defined and symmetry-aware architectures. Its role is not to replace the geometry-generation layer, but to sit directly on top of it. In the present stack, GeoCYData provides the geometric substrate---case definitions, bundle generation, local and invariant feature views, symmetry-aware metadata, and benchmark protocols---while GlobalCY provides the learned model families, metric construction, diagnostic evaluation, comparison workflows, and result-export layer. This separation of concerns is deliberate: it allows the geometry layer to remain fixed while the learned-model layer evolves, and it ensures that architectural comparisons are not confounded by changes in the underlying geometric data. This emphasis is intentionally different from algebraic-ansatz approaches in which compatibility properties are built directly into the parametrization itself; here the goal is to test how far globally structured learned representations alone improve hard-regime behavior under a common geometric evaluation pipeline \cite{GerdesKrippendorf2022}.

The framework is organized around a small number of research-critical layers. At the input layer, GlobalCY consumes GeoCYData bundles and converts them into model-ready representations. At the model layer, it supports three learned K\"ahler-potential families: a local-input baseline, a globally invariant model, and a symmetry-aware global model. At the geometric layer, it constructs Hermitian metric corrections from the learned scalar \(\phi\) using JAX automatic differentiation. At the diagnostic layer, it computes eigenvalue summaries, negativity-related quantities, projective-invariance drift, symmetry consistency, and related comparison statistics. At the experiment layer, it supports smoke-scale validation runs, multi-seed ablations, aggregated comparisons, and a lightweight result-export workflow for manuscript-facing tables and figures.

The framework is intentionally focused. It is not designed, at this stage, as a universal package for all learned Calabi--Yau metric experiments. Rather, it is built around a specific scientific question: whether global invariant structure and symmetry-aware structure improve learned K\"ahler-potential behavior on hard quartic regimes. The software architecture reflects that goal directly. Each layer corresponds to a research need: ingest the appropriate geometric views, train the relevant model classes, evaluate geometry-sensitive diagnostics, and export stable comparison artifacts for scientific interpretation. 

\begin{figure}[tbp]
\centering
\includegraphics[width=\textwidth]{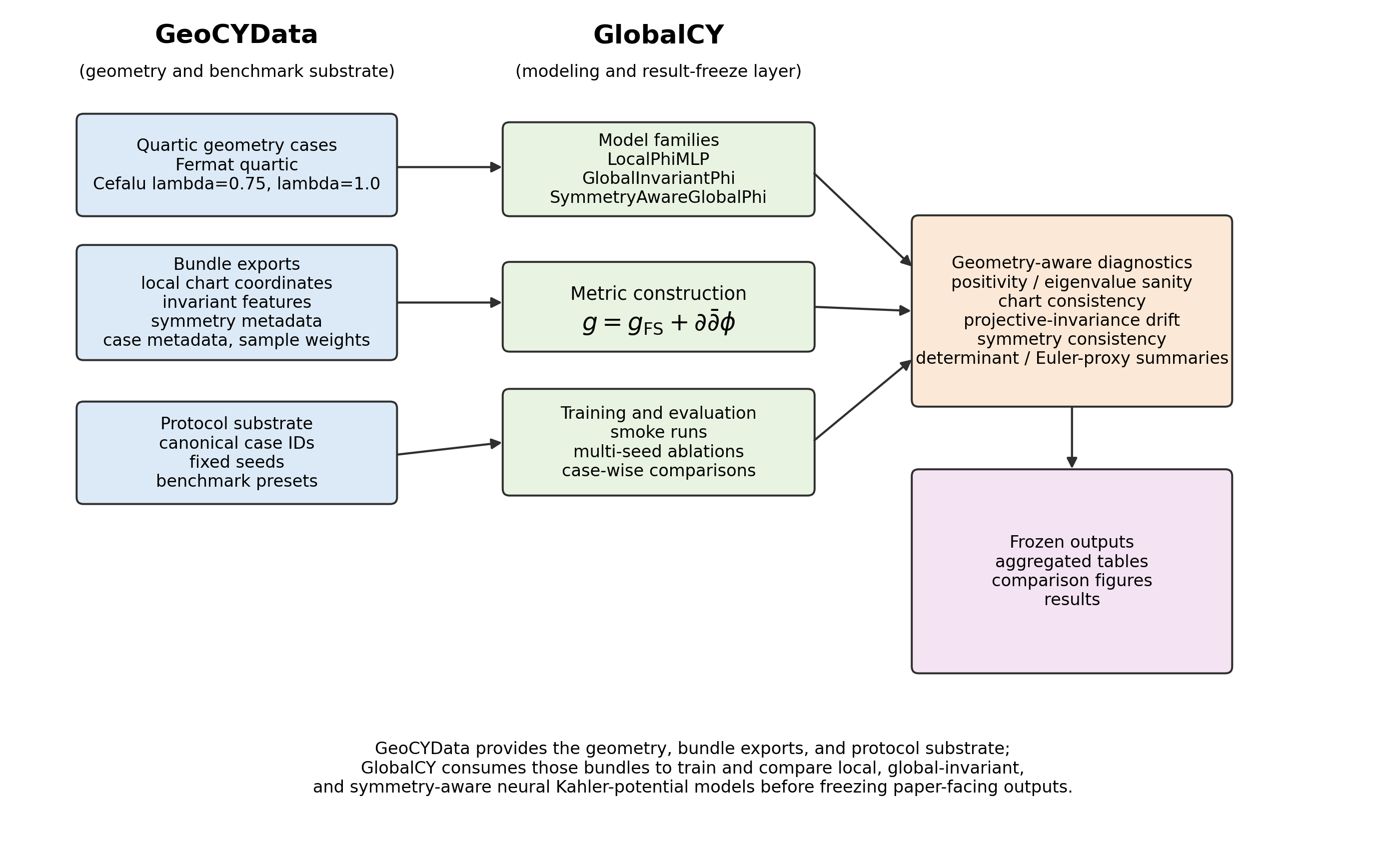}
\caption{High-level architecture of the GlobalCY I benchmark stack. The workflow is organized around a two-layer division of responsibilities. \texttt{GeoCYData} provides the geometric and protocol substrate, including quartic geometry cases, bundle exports with local-chart, invariant, and symmetry-aware views, and the canonical case identifiers, seeds, and benchmark presets used for controlled comparison. \texttt{GlobalCY} consumes those bundles to instantiate the three model families studied here---\texttt{LocalPhiMLP}, \texttt{GlobalInvariantPhi}, and \texttt{SymmetryAwareGlobalPhi}---construct learned metric corrections through the K\"ahler-potential ansatz \(g = g_{\mathrm{FS}} + \partial \bar{\partial}\phi\), and evaluate them with geometry-aware diagnostics. The same stack supports smoke validation, multi-seed ablations, case-wise aggregation, and frozen paper-facing outputs such as comparison tables, figures, and summary artifacts. The figure highlights that the present study is not only a model comparison, but a reproducible geometry-to-results pipeline in which the geometric substrate is held fixed while architectural assumptions at the learned-model layer are varied in a controlled way.}
\label{fig:globalcy-architecture}
\end{figure}

The main architectural point of Figure~\ref{fig:globalcy-architecture} is that the comparison is controlled at the level of the geometric substrate. The local, globally invariant, and symmetry-aware models are not trained on unrelated datasets or under unrelated preprocessing conventions; rather, they are trained on different model-facing views of the same underlying quartic cases. This separation makes it easier to attribute observed differences in performance to architectural structure rather than to changes in geometry generation or experiment assembly.

\subsection{Bundle interface with GeoCYData}

GlobalCY consumes model-facing bundles exported by GeoCYData. These bundles provide multiple coordinated views of the same underlying geometric case, allowing architectural comparisons to be carried out without regenerating geometry separately for each model family. The basic view consists of local chart coordinates, which support the local-input baseline. A second view consists of projective-invariant feature representations, which support the globally invariant model. A third view, when available, includes symmetry-aware information such as canonical representatives, canonical invariant features, orbit metadata, and symmetry reference views. The bundle interface also carries explicit case metadata, including geometry labels, case identifiers, parameter values, and seed information, so that experiments can be reproduced exactly and aggregated coherently.

This bundle design is one of the main reasons the architectural comparison is controlled. The local, global, and symmetry-aware models are not trained on unrelated datasets or under unrelated preprocessing conventions. They are trained on different representations of the same geometric cases, all produced from a single benchmark substrate. This avoids a common failure mode in scientific machine learning, namely that apparent architectural improvements are actually caused by differences in preprocessing or data construction. In the present framework, the distinction between models is intended to reflect a difference in architectural assumptions rather than a difference in what geometry they were given.

In practice, the bundle-adapter layer serves as the translation point between GeoCYData outputs and GlobalCY inputs. It exposes chart-local features to the local baseline, invariant feature tensors to the global model, and canonicalized or orbit-aware representations to the symmetry-aware model. It also retains the metadata needed for per-case and multi-seed reporting. GeoCYData and GlobalCY therefore form a two-layer stack: the former provides the geometric substrate, while the latter provides the learned metric and diagnostic layer.

\subsection{Model families}

\subsubsection{LocalPhiMLP}

The first model family is a local-input baseline, denoted \texttt{LocalPhiMLP}. This model consumes chart-local coordinate data and learns a scalar correction \(\phi\) using a standard multilayer perceptron. It is the most direct neural analog of a patch-local \(\Phi\)-style ansatz: expressive enough to fit local structure, but not explicitly constrained to respect global geometric structure beyond what is learned implicitly from the data. It therefore provides a natural baseline against which to test whether stronger architectural structure improves geometric behavior.

This baseline is important for two reasons. First, it reflects a modeling strategy that is straightforward to implement and has served as a natural starting point in related learned-metric work. Second, it exposes the very weakness that motivates the present paper: even when such a model fits local behavior reasonably well, it may fail to act like a globally meaningful K\"ahler-potential correction when judged by projective, symmetry-sensitive, or topology-sensitive criteria. The role of \texttt{LocalPhiMLP} is therefore not to represent an ideal endpoint, but to provide a concrete local-input reference against which more structured alternatives can be assessed.

\subsubsection{GlobalInvariantPhi}

The second model family is the globally defined invariant model, denoted \texttt{GlobalInvariantPhi}. Instead of consuming chart-local coordinates, this model is trained on projective-invariant feature representations derived from the ambient hypersurface data. The goal is to make the learned scalar correction \(\phi\) depend on features that are already compatible with the global projective structure of the underlying geometry. In this sense, \texttt{GlobalInvariantPhi} is not merely a larger or smaller network than the local baseline; it is a different ansatz class, one in which the input representation itself carries part of the desired geometric constraint.

This is the central architectural object of the present paper. The underlying idea is simple: if the learned correction is intended to behave like a global geometric scalar, then its inputs should not remain inherently patch-local when a global invariant representation is available. The empirical comparison suggests that this shift is meaningful. The globally invariant model is strongest overall on the clearest geometry-sensitive metrics in the benchmark, indicating that global invariant structure already provides a real improvement over the purely local-input baseline.

\subsubsection{SymmetryAwareGlobalPhi}

The third model family is the symmetry-aware global model, denoted \texttt{SymmetryAwareGlobalPhi}. Like the invariant global model, it is built from globally meaningful feature representations rather than raw chart-local inputs. In addition, it incorporates symmetry-aware information supplied by GeoCYData, including canonical invariant views and orbit-related context. In the current implementation, this symmetry-aware structure is deliberately modest and inspectable: it introduces a small symmetry-conditioned modification to the global invariant architecture rather than a fully equivariant or group-theoretically elaborate construction.

This design is intentional. The purpose of the present study is to test whether symmetry-aware structure helps in a first controlled comparison, not to claim that the current implementation is the strongest possible realization of the idea. The model should therefore be understood as a first operational symmetry-aware architecture rather than an endpoint. It is sufficiently nontrivial to support a meaningful symmetry-consistency diagnostic and to improve over the local baseline on selected metrics, especially projective-invariance drift, but it does not yet surpass the plain globally invariant model in the current comparison. That outcome is still scientifically informative: it suggests that symmetry-awareness is promising, but that in its current form it is not yet the dominant architectural ingredient.

\subsection{Metric construction}

All three model families are interpreted through the same K\"ahler-potential correction ansatz
\[
g = g_{\mathrm{FS}} + \partial \bar{\partial}\phi,
\]
where \(g_{\mathrm{FS}}\) is the reference Fubini--Study metric and \(\phi\) is the learned scalar correction. GlobalCY uses JAX automatic differentiation to compute the derivatives required for this construction, producing Hermitian metric matrices from the learned scalar field in a fully differentiable and reproducible way. This is one of the main reasons the framework is JAX-native from the outset: model evaluation and metric construction are carried out in the same computational environment rather than coupled only afterwards through offline postprocessing.

The resulting metric objects are summarized by a collection of geometry-sensitive statistics, including determinant summaries, eigenvalue summaries, negativity-related quantities, and projective- and symmetry-sensitive diagnostics defined at the experiment layer. The point of these summaries is not to claim that every relevant geometric invariant has already been captured, but to ensure that the learned \(\phi\)-models are evaluated as metric corrections rather than merely as abstract regressors.

A practical advantage of this design is that the metric-construction layer is shared across all three model families. This sharpens the comparison: differences in the reported results cannot be attributed to one model being postprocessed through a different geometric pipeline than another. They arise from differences in the learned scalar correction under a common geometric construction. In this sense, the metric layer acts as a geometric equalizer: all three models enter the same correction ansatz, but with different architectural assumptions about what \(\phi\) depends on and how much global structure is built in from the start.
\section{Benchmarking and Test Harness}

\subsection{Why a controlled harness is needed}

GeoCYData and GlobalCY together define a controlled computational harness in which geometry generation, model comparison, diagnostic evaluation, and result export are reproducible under a shared case protocol. This is essential because the scientific question is not whether a single model can be trained in isolation, but whether distinct architectural constraints can be compared fairly on the same hard quartic regimes. In learned Calabi--Yau metric approximation, even modest differences in case definition, preprocessing, or evaluation logic can blur the architectural effect one is trying to isolate. The purpose of the present harness is therefore to hold the geometric substrate fixed while allowing the model class to vary in a controlled and interpretable way.

The harness is naturally two-layered. GeoCYData provides the geometry and protocol side: quartic-family case definitions, bundle generation, local and invariant feature views, symmetry-aware metadata, case identifiers, and benchmark presets. GlobalCY provides the learned-model side: local, global, and symmetry-aware K\"ahler-potential models; geometry-aware diagnostics; multi-seed comparison workflows; and the result-export layer used for manuscript construction. This division is useful computationally because it prevents geometry generation and model comparison from becoming entangled, and useful scientifically because it makes observed differences in model behavior easier to attribute to architecture rather than to experiment assembly.

A controlled harness is particularly important in hard quartic regimes. In easier settings, several model classes may appear superficially acceptable, making it difficult to determine whether a more structured ansatz is genuinely better or merely more elaborate. The Cefal\'u cases considered here serve instead as stress tests. In such a regime, uncontrolled variation in bundle construction, seed choice, or reporting format would weaken the comparison substantially. The shared harness ensures that the central comparison is really about local versus global versus symmetry-aware structure.

\subsection{Comparison protocol}

The benchmark considered in this paper is deliberately narrow. We compare three model families on two hard Cefal\'u cases using a fixed three-seed protocol. The geometric cases are
\[
\lambda = 0.75 \qquad \text{and} \qquad \lambda = 1.0,
\]
and the model families are
\begin{itemize}[leftmargin=2em]
    \item \texttt{local},
    \item \texttt{global},
    \item \texttt{symmetry\_aware}.
\end{itemize}
For each case, the models are evaluated across the fixed seed set
\[
7,\;11,\;19,
\]
and the per-seed outputs are aggregated into a multi-seed comparison layer. The resulting study is modest in scale but strong in interpretability: it is small enough to inspect closely, yet large enough to avoid overinterpreting a single initialization.

The comparison proceeds in three stages. First, GeoCYData bundles are generated for the chosen Cefal\'u cases with the model-facing views required by the three architectures. Second, GlobalCY runs per-case ablations across the selected seeds and records per-run diagnostics together with machine-readable comparison artifacts. Third, the \texttt{freeze\_results} workflow consumes the aggregated multi-seed outputs and writes a stable result package containing tables, figures, and summary files. In this way, the protocol is not merely a training workflow, but an end-to-end scientific comparison workflow.

\begin{table}[H]
\centering
\caption{Controlled comparison protocol used in the benchmark. Two hard Cefal\'u cases, three model families, and a fixed three-seed evaluation scheme are used throughout.}
\label{tab:paper1-protocol}
\begin{tabular}{@{}ll@{}}
\toprule
Component & Setting \\
\midrule
Geometry family & \texttt{cefalu\_quartic} \\
Cases & \texttt{cefalu\_lambda\_0\_75}, \texttt{cefalu\_lambda\_1\_0} \\
Model families & \texttt{local}, \texttt{global}, \texttt{symmetry\_aware} \\
Seed set & \(7,11,19\) \\
Per-case workflow & multi-seed ablation + comparison aggregation \\
Result-export workflow & \texttt{freeze\_results} \\
Primary exported artifacts & tables, figures, JSON, summary memo \\
\bottomrule
\end{tabular}
\end{table}

\subsection{Diagnostics}

We evaluate model families using a geometry-aware diagnostic suite designed to measure not only optimization quality but also the extent to which a learned K\"ahler-potential correction behaves like a geometrically meaningful object. The full set of diagnostics is summarized in Table~\ref{tab:paper1-diagnostics}. It includes measures of negative-eigenvalue behavior, lower-tail spectral behavior, chart consistency, projective-invariance drift, symmetry consistency, determinant- and Euler-proxy summaries, training loss, and runtime. This choice of diagnostics is intentionally more geometric than a standard machine-learning benchmark. Training loss is retained because optimization quality still matters, but it is not treated as the primary scientific criterion. Instead, the greatest interpretive weight is assigned to diagnostics that most directly test whether the learned correction remains stable and globally meaningful on hard quartic regimes.

Among these quantities, two play the central role in the present comparison. The first is the negative-eigenvalue frequency, which measures how often the learned metric correction exhibits undesirable negative-eigenvalue behavior and therefore serves as a basic sanity and stability diagnostic. The second is projective-invariance drift, which probes whether the learned model respects the global projective structure that motivates the invariant ansatz. Taken together, these quantities provide the clearest test of the main architectural claim of the paper: that globally structured learned representations should behave more faithfully than local-input baselines in hard regimes where geometric fragility is already known to appear.

The remaining diagnostics refine and qualify this picture. The minimum-eigenvalue summary tracks the lower tail of the Hermitian spectrum and complements the negativity rate. Chart consistency is included to assess whether the learned corrections remain well behaved across local-coordinate descriptions, although in the present benchmark it is not yet strongly discriminative. Symmetry consistency probes whether the symmetry-aware model family yields gains beyond purely local or purely invariant modeling, though its interpretation must remain cautious because the current symmetry-aware architecture is intentionally modest. Finally, determinant and Euler-proxy summaries serve as supporting geometry-facing checks that keep the benchmark from collapsing into a purely optimization-driven exercise, while runtime records the operational cost of the comparison workflow. The central conclusions of the paper are therefore drawn from the strongest architectural signal visible in negativity behavior and projective drift, with the remaining quantities serving to qualify, limit, and contextualize that signal.

\begin{table}[H]
\centering
\caption{Diagnostic suite used in the benchmark. The final column indicates the main interpretive role of each diagnostic in the current comparison.}
\label{tab:paper1-diagnostics}
\setlength{\tabcolsep}{6pt}
\renewcommand{\arraystretch}{1.08}
\begin{tabular}{@{}>{\raggedright\arraybackslash}p{3.1cm} >{\raggedright\arraybackslash}p{7.3cm} >{\raggedright\arraybackslash}p{2.8cm}@{}}
\toprule
Diagnostic & Interpretation & Role \\
\midrule
\texttt{Negative fraction}
& Frequency of negative-eigenvalue behavior in the learned metric correction
& Primary \\
\midrule
\texttt{Minimum eigenvalue mean}
& Mean lower-tail eigenvalue behavior
& Secondary \\
\midrule
\texttt{Chart consistency}
& Chart-overlap consistency summary
& Secondary / nondiscriminating in current benchmark \\
\midrule
\texttt{Projective invariance drift}
& Drift under projective-invariant comparison; key global-structure diagnostic
& Primary \\
\midrule
\texttt{Symmetry consistency}
& Symmetry-sensitive comparison summary
& Secondary / architecture-sensitive \\
\midrule
\texttt{Determinant mean}
& Mean determinant summary for the learned metric correction
& Auxiliary \\
\midrule
\texttt{Euler proxy}
& Lightweight topology-facing proxy summary
& Auxiliary \\
\midrule
\texttt{Training loss}
& Optimization objective summary
& Supporting only \\
\midrule
\texttt{Runtime seconds}
& Computational cost summary
& Operational \\
\bottomrule
\end{tabular}
\renewcommand{\arraystretch}{1.0}
\end{table}

\subsection{Result-freeze workflow}

To convert the benchmark into a stable paper-writing workflow, GlobalCY includes a lightweight result-freeze layer. The \texttt{freeze\_results} command consumes the existing multi-seed comparison directories for the selected Cefal\'u cases rather than rerunning the benchmark itself. It then writes a compact output directory containing manuscript-facing tables, figures, machine-readable summaries, and a short results memo. This separation is practically important: once the comparison runs have been completed, the manuscript can be written against a fixed set of artifacts rather than against transient experimental outputs that may continue to evolve.

In the present workflow, the result-freeze layer combines the comparison outputs for the two retained Cefal\'u benchmark cases and produces a compact set of manuscript-facing artifacts. These include machine-readable comparison tables, readable summary files, a consolidated JSON result package, exported comparison figures, and a short drafting memo. Together, these artifacts are sufficient to support the Results and Discussion sections without requiring the full experiment graph to be rerun during manuscript preparation.

\begin{table}[tbp]
\centering
\caption{Result artifacts produced by the \texttt{freeze\_results} workflow.}
\label{tab:paper1-frozen-artifacts}
\setlength{\tabcolsep}{6pt}
\renewcommand{\arraystretch}{1.08}
\begin{tabular}{@{}>{\raggedright\arraybackslash}p{5.2cm}>{\raggedright\arraybackslash}p{8.0cm}@{}}
\toprule
Artifact & Role in manuscript workflow \\
\midrule
\texttt{paper1\_core\_results.csv} & machine-readable core comparison table \\
\texttt{paper1\_core\_results.md} & readable core comparison summary \\
\texttt{paper1\_robustness.csv} & machine-readable robustness/interpretation table \\
\texttt{paper1\_robustness.md} & readable robustness/interpretation summary \\
\texttt{paper1\_results.json} & consolidated machine-readable result package \\
\texttt{fig\_core\_comparison.png} & core comparison figure across selected cases and models \\
\texttt{fig\_hardest\_case.png} & hardest-case figure focused on \(\lambda = 1.0\) \\
\texttt{paper1\_summary.md} & compact narrative memo for drafting support \\
\bottomrule
\end{tabular}
\renewcommand{\arraystretch}{1.0}
\end{table}

Taken together, the benchmarking harness and result-freeze workflow provide the study with a controlled experimental backbone. They ensure that the reported findings are generated from a fixed geometric substrate, compared under a stable protocol, and exported in a form suitable for direct interpretation and manuscript construction.

\section{Experimental Setup}

The experimental design is intentionally narrow and controlled. Rather than attempting a broad first-pass survey across many quartic families or parameter values, we focus on two hard cases from the Cefal\'u family, \(\lambda = 0.75\) and \(\lambda = 1.0\). On each case, we compare three model families---a local-input baseline, a globally invariant model, and a symmetry-aware global model---under a fixed three-seed protocol using seeds \(7\), \(11\), and \(19\). The benchmark therefore comprises \(18\) core runs in total. This scale is deliberately modest: it is sufficient to support a meaningful architectural comparison while remaining small enough that the geometric interpretation of the resulting behavior stays clear.

The workflow is organized in three stages: smoke validation, full multi-seed comparison, and result export. Smoke-scale runs are used first to validate the model and data paths on the selected cases. The full comparison is then executed across all three model families and all three seeds for each case, after which per-seed outputs are aggregated into case-level summaries. Finally, the \texttt{freeze\_results} workflow produces the stable artifact set used in the manuscript. This layered design separates software validation from scientific aggregation and then separates both from manuscript-facing production.

Within the full diagnostic suite, two quantities carry the greatest interpretive weight: negative-eigenvalue frequency and projective-invariance drift. These are the primary comparison metrics because they most directly test the architectural claim at the center of the paper. Negative-eigenvalue frequency measures how often the learned metric correction exhibits undesirable negative-eigenvalue behavior and thus functions as a basic stability diagnostic. Projective-invariance drift probes whether the learned model respects the global projective structure that motivates the invariant ansatz. The remaining diagnostics are still informative, but they serve mainly secondary or qualifying roles in the interpretation of the benchmark.

\section{Results}

\subsection{Core comparison summary}

The benchmark yields a clear overall pattern. Across both hard Cefal\'u cases, the globally defined invariant model is the strongest overall model family on the two geometry-sensitive diagnostics that carry the greatest interpretive weight in this study: negative-eigenvalue frequency and projective-invariance drift. In both cases, the global model improves on the local baseline by reducing the frequency of negative-eigenvalue behavior and by substantially lowering projective-invariance drift. It also achieves lower mean training loss in both regimes. These gains are strongest at \(\lambda = 0.75\), while the \(\lambda = 1.0\) case is more difficult in the sense that the separation between the global and local models is smaller there.

The symmetry-aware model occupies an intermediate position in the current benchmark. It preserves part of the projective-drift improvement relative to the local baseline in both cases, but it does not outperform the plain globally invariant model on the main tracked metrics. In particular, it trails the global model on negative-eigenvalue frequency, projective-invariance drift, and training loss in both regimes, and it exhibits greater seed-to-seed variability on several quantities. The main positive result of the paper is therefore not that symmetry-awareness already dominates, but that global invariant structure provides a measurable and reproducible improvement over local-input modeling on this hard quartic benchmark.

\begin{table}[H]
\centering
\caption{Core comparison across the two hard Cefal\'u cases. Lower values are better for negativity, projective-invariance drift, and training loss.}
\label{tab:core-results}
\begin{tabular}{@{}llccc@{}}
\toprule
Case & Model & Negativity & Drift & Train loss \\
\midrule
\multirow{3}{*}{Cefal\'u $\lambda=0.75$}
& Local            & 0.08854 & \(4.39\times 10^{-8}\) & 11.94 \\
& Global           & 0.04167 & \(1.44\times 10^{-8}\) & 11.02 \\
& Symmetry-aware   & 0.12500 & \(2.17\times 10^{-8}\) & 12.43 \\
\midrule
\multirow{3}{*}{Cefal\'u $\lambda=1.0$}
& Local            & 0.04688 & \(3.95\times 10^{-8}\) & 10.65 \\
& Global           & 0.04167 & \(1.57\times 10^{-8}\) & 9.73 \\
& Symmetry-aware   & 0.13021 & \(2.02\times 10^{-8}\) & 11.44 \\
\bottomrule
\end{tabular}
\end{table}

\begin{table}[H]
\centering
\caption{Compact interpretation summary. The globally invariant model is strongest overall, with the clearest gains at \(\lambda=0.75\).}
\label{tab:robustness}
\begin{tabular}{@{}p{3.2cm}p{3.3cm}p{5.2cm}p{3.1cm}@{}}
\toprule
Case & Strongest model & Strongest gains & Hardness note \\
\midrule
Cefal\'u $\lambda=0.75$
& Global
& Large improvement over local in negativity and projective drift
& Clearer separation between global and local \\
\midrule
Cefal\'u $\lambda=1.0$
& Global
& Improvement over local remains, but the negativity gain is smaller
& Harder case \\
\bottomrule
\end{tabular}
\end{table}

\begin{figure}[tbp]
\centering
\includegraphics[width=0.78\textwidth]{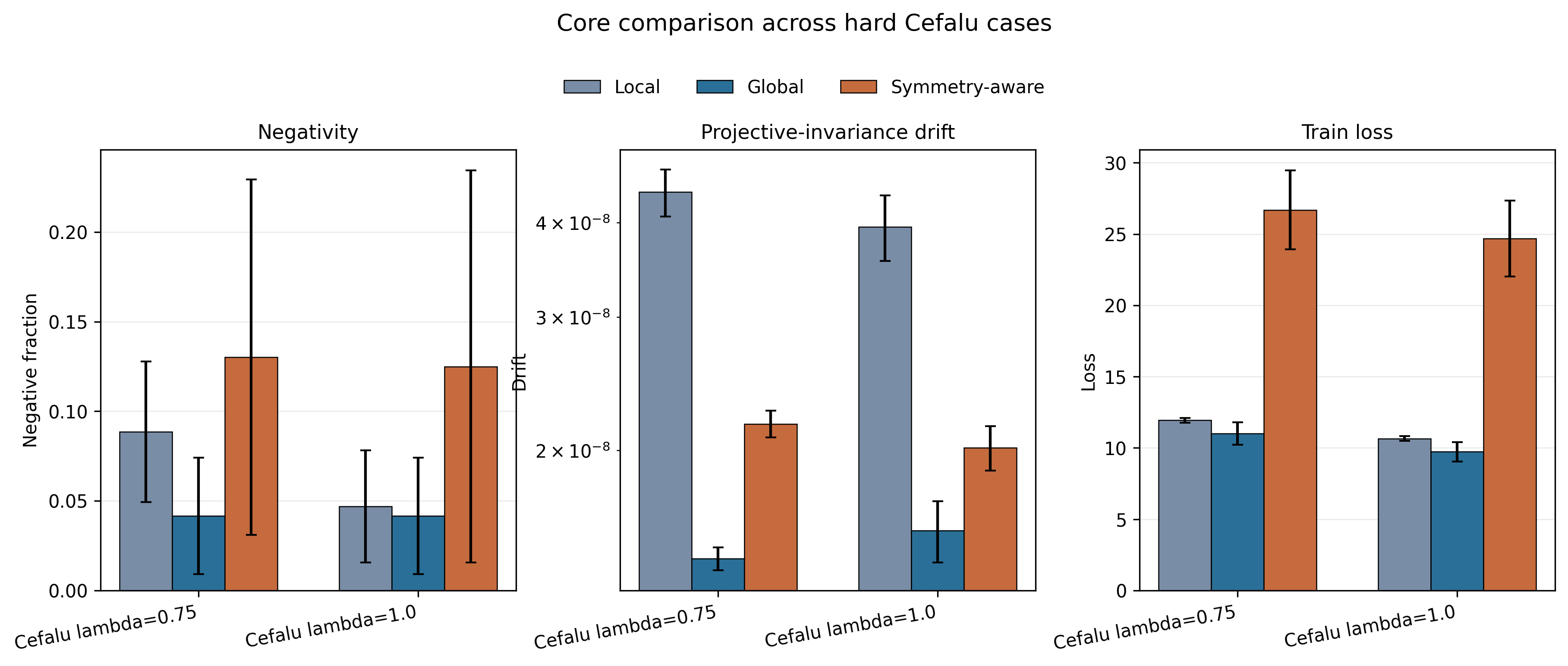}
\caption{Benchmark-wide comparison across the two hard Cefal\'u cases, \(\lambda=0.75\) and \(\lambda=1.0\), for the local, globally invariant, and symmetry-aware model families. The three panels report mean negative-eigenvalue frequency, mean projective-invariance drift, and mean training loss, with error bars indicating cross-seed variation over the fixed three-seed protocol. The figure shows that the globally invariant model is the strongest overall architecture in this benchmark: it achieves the lowest projective-invariance drift and the lowest training loss in both cases, and it also yields the lowest negative-eigenvalue frequency among the compared models. The separation between the global and local baselines is more pronounced at \(\lambda=0.75\) than at \(\lambda=1.0\), indicating that \(\lambda=1.0\) is the more difficult regime. The symmetry-aware model preserves part of the drift improvement relative to the local baseline, but its larger variability and weaker negativity behavior show that the present symmetry-aware implementation does not yet surpass the plain globally invariant architecture.}
\label{fig:frozen-core-comparison}
\end{figure}

\begin{figure}[tbp]
\centering
\includegraphics[width=0.78\textwidth]{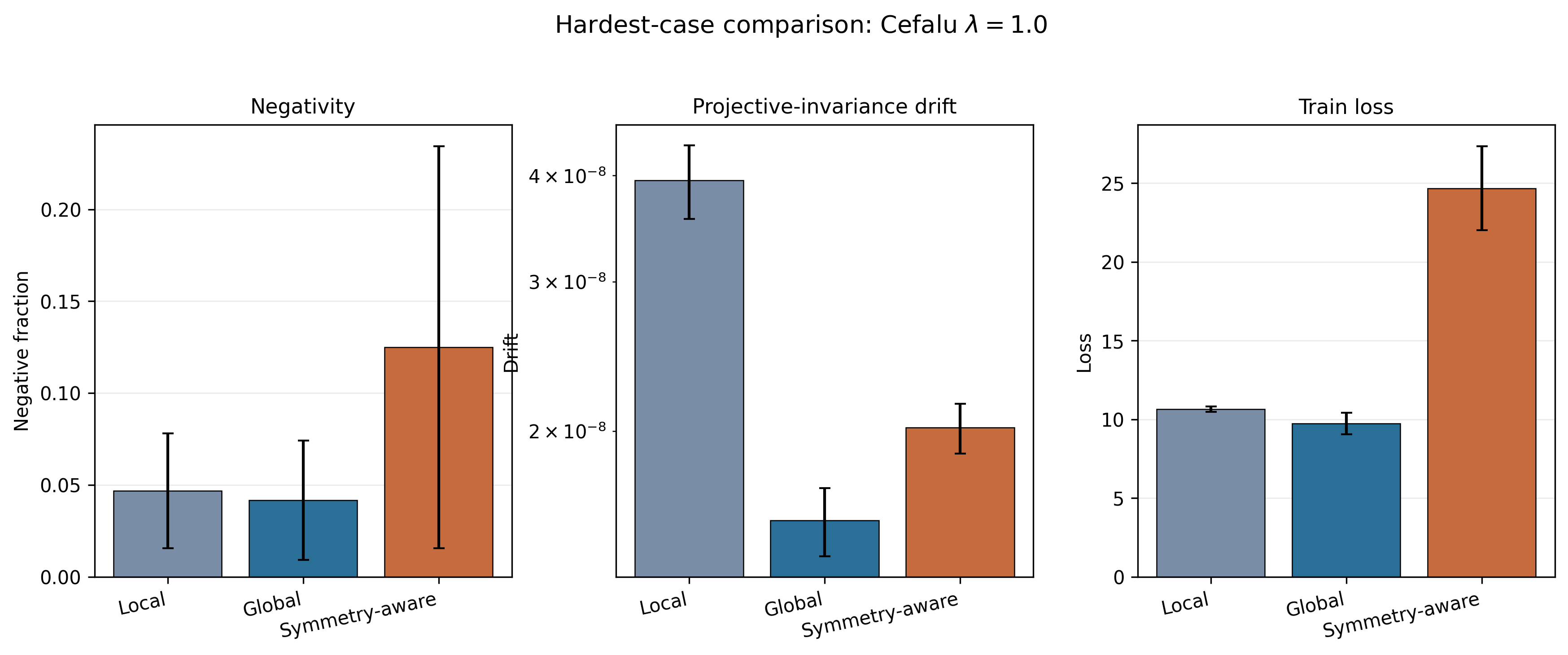}
\caption{Focused comparison for the hardest benchmark case, the Cefal\'u regime \(\lambda=1.0\). The three panels isolate the same diagnostics as in Figure~\ref{fig:frozen-core-comparison}: mean negative-eigenvalue frequency, mean projective-invariance drift, and mean training loss, again with error bars indicating cross-seed variation. This figure clarifies why \(\lambda=1.0\) is interpreted as the harder regime. The globally invariant model still outperforms the local baseline, particularly in projective-invariance drift and training loss, but the negativity improvement is smaller than in the \(\lambda=0.75\) case. The symmetry-aware model continues to reduce drift relative to the local baseline, but its higher variability and worse negativity rate indicate that the current symmetry-aware architecture is not yet as well conditioned as the simpler globally invariant model.}
\label{fig:frozen-hardest-case}
\end{figure}

\subsection{Case-by-case findings}

The \(\lambda=0.75\) Cefal\'u case exhibits the clearest improvement from local to global invariant modeling. The global model reduces the mean negativity rate from \(0.08854\) to \(0.04167\), while simultaneously reducing projective-invariance drift from \(4.39\times 10^{-8}\) to \(1.44\times 10^{-8}\). It also lowers the mean training loss from \(11.94\) to \(11.02\). In this case, the symmetry-aware model improves the drift relative to the local baseline, reaching \(2.17\times 10^{-8}\), but it does not match the global model on negativity or training loss. It also shows the largest across-seed variability on several metrics, including negativity, training loss, and Euler-proxy summaries.

The \(\lambda=1.0\) case tells the same qualitative story, but in a more difficult regime. The global model again outperforms the local baseline on the two principal comparison metrics, reducing the negativity rate from \(0.04688\) to \(0.04167\) and reducing projective-invariance drift from \(3.95\times 10^{-8}\) to \(1.57\times 10^{-8}\). It also lowers the mean training loss from \(10.65\) to \(9.73\). The symmetry-aware model again improves drift relative to the local baseline, reaching \(2.02\times 10^{-8}\), but it remains weaker than the global model on the main metrics and again exhibits larger variability across seeds. Thus, while the global model remains strongest overall at \(\lambda=1.0\), the margin of superiority is smaller than at \(\lambda=0.75\), which is why this regime is interpreted as the harder case.

One important nuance is that the local model retains the strongest mean \texttt{min\_eigenvalue\_mean} in both cases. This shows that the architectural gains of the global model are not uniform across all diagnostics. The claim of the paper is therefore not that global invariant models dominate every metric simultaneously, but that they dominate on the principal geometry-sensitive comparison metrics most directly tied to the central thesis, namely negative-eigenvalue frequency and projective-invariance drift.

\subsection{Cross-case interpretation}

Taken together, the results support three main conclusions. First, the globally defined invariant model is the strongest overall model family in the current benchmark. It outperforms the local baseline on \texttt{negative\_fraction}, \texttt{projective\_invariance\_drift}, and \texttt{train\_loss} in both hard Cefal\'u cases. Second, the strongest gains occur at \(\lambda=0.75\), where the separation between global and local models is clearest. Third, the \(\lambda=1.0\) regime is harder, since the improvement of the global model over the local baseline is smaller there, especially on the negativity metric.

The symmetry-aware model leads to a more nuanced conclusion. It preserves part of the global-structure advantage over the local baseline, especially in the reduction of projective-invariance drift. However, in its current modest form it does not outperform the plain global invariant model on the main diagnostics, and it is the most unstable model family across seeds. This suggests that symmetry-awareness is promising as an architectural direction, but that the present implementation is not yet strong enough to justify a claim of superiority over the simpler invariant global model.

A final interpretive point concerns what \emph{does not} currently separate the models. The reported \texttt{chart\_consistency} values are \(0.0\) for all rows, so this metric does not distinguish the architectures in the present benchmark. Likewise, the apparent \texttt{symmetry\_consistency} advantage of the local baseline should not be interpreted literally, since the local path reports a degenerate value there by construction rather than through a meaningful symmetry-aware evaluation. These observations further justify centering the paper's main claims on the diagnostics that do exhibit a robust architectural signal.

\section{Discussion}

\subsection{What the results establish}

The results support a clear but deliberately bounded conclusion: globally defined invariant structure improves the geometric behavior of learned K\"ahler-potential corrections on the hard quartic benchmark studied here. This conclusion is grounded in the two diagnostics that most directly test the paper's central thesis, namely \texttt{negative\_fraction} and \texttt{projective\_invariance\_drift}. Across both Cefal\'u cases, the globally invariant model reduces the frequency of negative-eigenvalue behavior and substantially lowers projective-invariance drift relative to the local baseline. These improvements persist across the fixed three-seed protocol and are therefore not artifacts of a single initialization.

This is scientifically meaningful because the main question of the paper is architectural rather than merely predictive. The issue is not whether one model attains a slightly lower loss than another in isolation, but whether supplying the learned ansatz with stronger global geometric structure produces a model that behaves more faithfully as a K\"ahler-potential correction in regimes where geometric fragility is already known to appear. On that question, the benchmark gives a positive answer. The gains are not uniform across every reported statistic, but they are clear on the geometry-sensitive quantities most tightly aligned with the stated hypothesis.

More broadly, the results show that the distinction between local and global learned K\"ahler-potential modeling is not cosmetic. In the present hard-regime comparison, that distinction is measurable. This is the main scientific value of the paper. It identifies a concrete architectural improvement that survives controlled comparison and therefore provides a principled starting point for subsequent work on stronger geometric diagnostics, more difficult singular regimes, and downstream computational applications.

At the same time, the evidence supports only a claim of this scope. The benchmark covers two hard Cefal\'u cases and a controlled family of model classes; it does not yet cover the full singular landscape of the family, nor does it establish a final geometry-certified learned metric pipeline. The paper should therefore be read as a first architectural result: it isolates one nontrivial improvement clearly, rather than claiming to solve the learned Ricci-flat metric problem in full generality.

\subsection{How to interpret the symmetry-aware results}

The symmetry-aware results are encouraging, but they do not yet support a claim of superiority. In both cases, the symmetry-aware model improves \texttt{projective\_invariance\_drift} relative to the local baseline, which indicates that incorporating symmetry-aware information into the learned ansatz is directionally sensible. However, it does not outperform the plain globally invariant model on the main diagnostics, and it exhibits larger variability across seeds on several quantities, including negativity behavior, training loss, and Euler-proxy summaries.

This should not be interpreted as evidence against symmetry-aware learning itself. Rather, it suggests that symmetry-awareness is not automatically beneficial unless it is paired with a sufficiently strong architecture or a sufficiently well-conditioned objective. The current \texttt{SymmetryAwareGlobalPhi} model is intentionally modest and inspectable. It is a first operational symmetry-aware construction, not a fully equivariant or deeply symmetry-adapted architecture. Its value in the present paper is therefore diagnostic: it demonstrates that a symmetry-aware lane is viable and nontrivial, while also showing that a first-layer implementation is not yet enough to surpass the simpler globally invariant model.

This is in fact a productive outcome for the broader program, because it separates two claims that could otherwise be blurred together. The first is that global invariant structure already yields a real gain over a local-input baseline. The second is that symmetry-aware structure appears promising, but requires further architectural development before a stronger claim can be justified. Keeping these claims distinct makes the present paper more credible and clarifies the next stage of the research program.

\subsection{Limitations and boundaries of the present study}

Several limitations remain visible in the current benchmark and should be stated plainly. The first is scope. The study is restricted to two hard Cefal\'u cases, \(\lambda=0.75\) and \(\lambda=1.0\). This is enough for a controlled first comparison, but not enough to support sweeping claims about all quartic regimes or all singular behaviors.

The second limitation is diagnostic resolution. The quantity \texttt{chart\_consistency} does not currently differentiate the model families, since it is \(0.0\) across all benchmark rows. This means that, at least in the present setup, it does not provide additional discriminative evidence. A third limitation is that the local baseline still achieves the strongest mean \texttt{min\_eigenvalue\_mean} in both cases. The architectural gains of the globally invariant model are therefore not universal across all reported summaries, and the paper should not be read as claiming otherwise.

A fourth limitation is stability. The symmetry-aware model is the most variable of the three model families across seeds, which indicates that the current implementation is not yet as well conditioned as the simpler invariant model. Finally, the present work still operates with a relatively lightweight geometry-facing diagnostic stack. The quantities used here are sufficient to support a meaningful first architectural conclusion, but they are not yet the full characteristic-form, curvature-rich, or observable-level pipeline that would be required for stronger claims in computational string phenomenology.

These limitations do not weaken the core result, but they do determine its correct interpretation. The present study should be viewed as establishing a first controlled architectural gain, not as delivering a final solution to learned Calabi--Yau metric computation or a fully developed phenomenology pipeline.

\subsection{Why this result matters for the broader program}

The importance of the present result lies not only in the specific benchmark numbers, but in what they enable conceptually. A central difficulty in this area is that many learned surrogates can appear acceptable on local or optimization-based criteria while failing to support stable geometric computation in the regimes that matter most. The present paper shows that imposing global invariant structure already changes that picture in a measurable way. This gives the broader program a firmer foundation: subsequent work can build on an ansatz class that has already demonstrated improved behavior on hard-regime diagnostics, rather than continuing to rely on purely local surrogates whose failure modes are already visible.

This matters especially for any future transition from architectural benchmarking to stronger scientific computation. Downstream quantities---whether curvature summaries, characteristic-form diagnostics, symbolic surrogates, moduli-dependent families, or more overtly physics-facing observables---will depend on the stability and geometric faithfulness of the learned metric layer beneath them. In that sense, the present paper addresses a prerequisite problem. It does not yet deliver the full downstream science pipeline, but it identifies a better starting architecture for building one.

\subsection{What follows from here}

The next steps suggested by the present work are reasonably clear. One immediate direction is to strengthen the training objective itself by incorporating additional geometry-aware constraints, for example penalties tied to chart consistency, positivity behavior, or other topology- and curvature-sensitive quantities. A second direction is to pursue symbolic distillation of the strongest model family. One of the long-term motivations for globally invariant modeling is precisely that it may yield learned K\"ahler-potential surrogates that are not only stronger empirically, but also more interpretable and more compressible.

A third direction is moduli-dependent learning. The present paper deliberately fixes attention on two hard benchmark points in order to isolate the architectural question cleanly. The next scientific step is to ask whether the improved architecture can be transported coherently across parameter families rather than retrained case by case. A fourth direction is singularity-aware modeling. The fact that the \(\lambda=1.0\) case remains harder strongly suggests that local asymptotics, blow-up-aware features, or more explicitly singularity-sensitive model classes may be needed in the more delicate regimes.

Finally, the longer-term direction is to connect the improved learned metric stack to richer geometry-sensitive and physics-facing computations. A particularly important next benchmark, not carried out here, is a direct comparison between the globally invariant learned models and algebraic-ansatz or Donaldson-style baselines on the same hard quartic regimes, in order to separate gains due to learned global structure from gains due to built-in algebraic compatibility. That includes more stable characteristic-form summaries, stronger curvature diagnostics, and eventually observable-level pipelines. In this sense, the present paper is best viewed as the first controlled architectural result in a larger program. It establishes that global invariant structure already provides a real advantage, shows that symmetry-aware modeling is promising but not yet mature, and identifies the harder regime in which future geometric and architectural refinements are most likely to matter.
\section{Conclusion}

This paper introduced \emph{GlobalCY}, a JAX-native framework for globally defined and symmetry-aware neural K\"ahler-potential models on projective hypersurface Calabi--Yau geometries. The goal was to address a specific architectural question: whether local-input learned K\"ahler-potential models can be improved by imposing stronger global geometric structure, particularly in hard quartic regimes where geometry-sensitive diagnostics are known to become fragile. To make that question testable, we combined GeoCYData as the geometry and protocol substrate with GlobalCY as the learned-model, diagnostic, and result-export layer, and compared three model families---local, globally invariant, and symmetry-aware global---on the Cefal\'u cases \(\lambda=0.75\) and \(\lambda=1.0\) under a fixed multi-seed protocol.

The main result is that the globally defined invariant model is the strongest overall model family in the benchmark considered here. Across both cases, it improves on the local baseline on the clearest geometry-sensitive comparison metrics, namely negative-eigenvalue frequency and projective-invariance drift, and it also achieves lower mean training loss. The gains are strongest at \(\lambda=0.75\), while the \(\lambda=1.0\) regime remains more difficult in the sense that the separation between global and local models is smaller there. These findings support the central claim of the paper: global invariant structure is already a scientifically meaningful architectural constraint for learned K\"ahler-potential modeling in hard quartic Calabi--Yau settings.

The paper also clarifies what cannot yet be claimed. The current symmetry-aware model is not strongest overall, and the architectural gains are not uniform across every diagnostic. In particular, the local baseline still attains the strongest mean minimum-eigenvalue statistic in both cases, while the present symmetry-aware model remains more variable across seeds and does not yet outperform the simpler globally invariant model. Nor does this study attempt to solve the full learned Ricci-flat metric problem, cover all quartic regimes, or provide a final observable-level computational pipeline. Its contribution is narrower and more precise: it establishes a first controlled architectural gain.

That result matters because it identifies a model-class improvement that survives controlled comparison in a hard geometric regime and therefore provides a stronger foundation for the next stage of the program. The distinction between local and global learned K\"ahler-potential modeling is not merely conceptual; in the present benchmark it is measurable. This, in turn, suggests a concrete roadmap for future work: stronger symmetry-aware architectures, richer geometry-aware training objectives, moduli-dependent extensions, singularity-aware model classes, and downstream geometry-sensitive or physics-facing computations. In that sense, this paper is not an endpoint, but a first rigorous step toward making hard Calabi--Yau metric geometry more computable, more interpretable, and more useful for broader geometric and string-theoretic applications.

%
%

\printbibliography


\end{document}